\documentclass[12pt]{article}
\pdfoutput=1
\usepackage{jheppub}

\usepackage{amsmath}
\usepackage{amssymb}
\usepackage{graphicx}

\usepackage{color}

\newcommand{\be}{\begin{equation}}
\newcommand{\ee}{\end{equation}}
\newcommand{\bea}{\begin{eqnarray}}
\newcommand{\eea}{\end{eqnarray}}
\newcommand{\nn}{\nonumber}

\title{Holographic Thermalization \\ in Quark Confining Background}

\author[a]{D.S. Ageev}
\author[a]{and I.Ya. Aref'eva}
\affiliation[a]{Steklov Mathematical Institute, RAS, Gubkin str. 8, 119991
Moscow, Russia}
\emailAdd{ageev@mi.ras.ru}
\emailAdd{arefeva@mi.ras.ru}

\abstract{We study holographic thermalization of a strongly coupled theory
inspired by  two colliding shock waves in a vacuum confining background.
Holographic thermalization means a black hole formation, in fact a trapped surface formation.
As a vacuum  confining background we considered  a well know bottom-up AdS/QCD model   that
provides the Cornell potential as well as reproduces QCD $\beta$-function. We perturb vacuum background
by  colliding domain shock waves, that are assumed to be holographically dual to heavy ions collisions.
Our main physical assumption is that we can make a restriction on
the time of a trapped surface formation that makes a natural limitation on the size of the domain where the trapped surface
is produced. This  limits the intermediate domain where the main part of the entropy
is produced.  In this  domain one can use an intermediate  vacuum background as an approximation
to the full confining background.
We have found that the dependence of multiplicity on energy for the intermediate  background  has an asymptotic 
expansion, which first term depends on energy as $E^{1/3}$, that is very similar to the experimental dependence of  particles multiplicities on  colliding ions energy obtained from RHIC and LHC. However, this first term, at the energies where 
the approximation of the confining metric by the intermediate  works,  does not saturate the exact answer
and one has to take into account the non-leading terms.  }


\begin{document}
\maketitle

\section{Introduction}

QCD, which is the currently accepted theory of strong interactions, still has the well-known problems with
description of a strong coupling  phenomena. The physics of heavy ion collisions, in particular QGP formation, involves real-time  strong coupled phenomena, that makes  difficult to study these phenomena within standard QCD methods.
 In the recent years a powerful  approach to QGP is explored. This method   is
based on
 a holographic duality between the strong coupling
quantum field in $d$-dimensional Minkowski space and classical
gravity in $d+1$-dimensional anti-de Sitter space (AdS)
\cite{Malda,GKP,Witten}. In particular, there is a considerable
progress in
 the holographic description of equilibrium  QGP  \cite{1101.0618}.
The holographic
 approach  is also applied to non-equilibrium  QGP.  Within this holographic
 approach
thermalization   is described  as a process of formation of a black
hole in AdS.

 The AdS/CFT correspondence is based on string theory and perfectly works for $\mathcal{N}$=4 SUSY  Yang Mills theory, while the dual description of real  QCD is unknown.
A lot of efforts have been  made  in searching for holographic QCD from string theory, in particular,
\cite{0306018,Mateos,SS}.  This approach is known as the "top-down" approach.
Other approach, known  as the
"bottom-up" approach,
is supposed to  propose a  suitable holographic
 QCD models from experimental data and lattice results \cite{HW,Andreev:2006ct,Karch2006,White:2007tu,Gursoy:2008za,Pirner:2009gr,He:2010ye,Gursoy:2010fj}.
 Main idea of this approach is using  natural prescriptions of the general AdS/CFT correspondence try to recover
 non-perturbative QCD phenomena, in particular  non-perturbative
 vacuum phenomena, finite temperature,  high-dense  and non-zero chemical potential phenomena.

   The 5-dim metrics that reproduce the Cornell potential \cite{Bali:2000gf}, as well as $\rho$-meson spectrum etc.,
   have been proposed \cite{Andreev:2006ct,Pirner:2009gr,He:2010ye}.  A so-called improved QCD (IHQCD) that is able to reproduce  the  QCD $\beta$-function
 has been constructed \cite{Gursoy:2010fj}.  Thermal deformation of these backgrounds are intensively studied in the last years, see for review \cite{1101.0618}.

 The problem of QGP formation is the subject of intensive study within holographic approach in last years, see
 \cite{IA,DeWolf} and references therein.
 There is a considerable progress in understanding of the thermalization process from the gravity side as BH formation.
 Initially this process has been considered starting from the AdS background  \cite{Gubser,0805.2927,Alv,0812.2053,0902.1508,Gubser2,ABG}.
However the pure AdS background is unable to
 describe  the vacuum QCD with quark confinement, as well it is not able to reproduce the QCD $\beta$-function.
 There are backgrounds that solve one, or even two of these problems. The first one has been solved in \cite{Andreev:2006ct}
 (see also \cite{Pirner:2009gr,He:2010ye}), where  a special version of soft-wall has been proposed, and the $\beta$-function  has been reproduced from IHQCD
 \cite{Gursoy:2008za,Gursoy:2010fj}.

 To describe the thermalization it is natural to study deformations of these backgrounds.
 Suitable deformations of  IHQCD by shock waves have been studied in \cite{KT, APP}  and it has been shown
 that {\it without} additional assumptions
IHQCD metric does not  reproduce the experimental multiplicity dependence on energy. In  \cite{Aref'eva:2014sua}  it has been noticed that holographic realization of the  experimental multiplicity
 requires an unstable background.

 The goal of this paper is to cover this gap and to show that the  model that reproduces
 the Cornell potential   at the same time  can be used as a gravity background to give a  correct energy dependence of multiplicities produced during a finite time.
 As a bonus of our approach we  get a reasonable estimation for the thermalization time.

 The paper is organized as follows. In Sect. 2.1  we remind confining metrics, that reproduce the Cornell potential.
 In Sect. 2.2 we remind the previous results concerning the multiplicities dependence on energy.
 In Sect. 2.3 we present the main formula for the size  of trapped surfaces formed in collision of the domain walls.
 In Sect. 3 we consider a special metric that is faraway from the confining metrics, but  gives
 a suitable entropy. We  also notice that a restriction of the size of the trapped surface permits
 to determine the thermalization time. In Sect. 4 we show that the confining metric \cite{He:2010ye} can be approximated at
 intermediate values of the holographic coordinate  $z$ by the
 metric  considered in Sect. 3. As a result this gives for the entropy produced during a short time, $\tau_{term}\sim 0.25fm$, an
 asymptotic expansion with the leading term which has  dependence on energy as $\sim E^{1/3}$. The same is true for metric \cite{Andreev:2006ct}.

\section{Setup}
\subsection{Confining Backgrounds}

It is well known that the AdS space does not reproduce the quark  confinement. To reproduce quark  confinement,
in particular the appropriate glueball spectrum,  Polchinski-Strassler \cite{HW} imposed the cut-off in the AdS space, "hard wall model".  Another modifications of the AdS space,  "soft wall models" \cite{Karch2006},  are related with
the  dilaton.
In the bottom-up approach, the metric is usually taken to be
\be
\label{metric-b}
ds^2=b^2(z)(-dt^2+dz^2+dx_i^2),
\ee
where $b^2(z)$ is some function usually taken to be the $AdS$ in the UV zone (this leads to the Coulomb potential in the UV) and is deformed $AdS$  in the IR. The deformation in the IR should be taken in such a way, that the quark-antiquark potential exhibits confinement.

 The experimental model of potential which is used to fit lattice and experimental data \cite{Bali:2000gf} is usually taken to be the Cornell potential. In principle this potential should reproduce quarkonia spectrum, interpolating between one-gluon exchange in the UV and linear confinement in the IR.

The model proposed  in \cite{Andreev:2006ct}  uses the following warp factor:
  \bea
 \label{b-AZ}
 b^2(z)&=&\frac{L^2h(z)}{z^2},\,\,\,\,\,
 h_{AZ}=e^{\frac{a z^2}{2}},\,\,\,a=0.42\,GeV^2.\eea
 In \cite{White:2007tu} it has been shown that this factor reproduces the static interquark potential obtained from $SU(3)$ lattice calculations \cite{Bali:2000gf}.

 In  \cite{He:2010ye} the  following modification
 \bea
 h(z)&=&
 \frac{\exp\left( -\frac{\sigma
 z^2}{2}\right)}{\left(\frac{z_{IR}-z}{z_{IR}}\right)^{c_0}},\,\,\,\,\sigma=0.34\,GeV^2, \,\,\,\,c_0=1,\,\,\,\,z_{IR}=2.54\,GeV^{-1}
 \eea
has been considered. This modification is in fact very close to
the model  \cite{Pirner:2009gr}  for $0 \,<z<2 \, fm$ and reproduces the Cornell potential and $\beta$-function.

In this paper we consider the modification (\ref{b2}), see below, of factor (\ref{b-AZ}), that also fits the Cornell potential well.
\subsection{Multiplicities}
  The experimental data for multiplicities in heavy-ion collisions at RHIC and LHC indicate
   \cite{ATLAS:2011ag}
    \be
    \label{M-exp}
   {\cal M}_{exp}\sim E^{0.3}+... \, .
   \ee
   Multiplicities obtained for the simplest holographic calculation in conformal background with the $AdS_5$ metric
  \cite{Gubser,0805.2927,Alv,0812.2053,0902.1508,Gubser2,ABG} ,
  \be
      \label{M-AdS}{\cal M}_{AdS_5}(E)\sim E^{2/3}
   \ee
 are  in fact worse than the Landau bound
   \be
       \label{M-Landau}
   {\cal M}_{Landau}(E)\sim E^{1/2}.
   \ee
To improve the energy dependence of  multiplicities    Kiritsis  and Taliotis \cite{KT}  have proposed  to use modifications
  of the $b$-factor.  They have considered $b$-factors corresponding to conformal  and non-conformal backgrounds.
  More precisely, they have considered the holographic point-like sources
  collision in dilaton models and got estimations for a variety of models (depending on the dilaton potential)
  \bea
      \label{M-Dilaton}{\cal M}_{a>1/3}&\sim &E^{\frac{3a+3}{3a+2}},\\
       \label{M-Dilaton-a}{\cal M}_{a\leq -1/3}&\sim &E^{\frac{3a+1}{3a}}.\eea
  Note that they also used  a perturbative QCD inspired   UV  cut-off.
  This modification provides $\log$-corrections.  Following  \cite{Gubser2}, where the energy-dependent cut-off in the high-energy limit has been proposed,   Kiritsis  and Taliotis \cite{KT} have shown that  this cut-off reduces  powers in (\ref{M-Dilaton}) and
  (\ref{M-Dilaton-a}) as
    \bea
      \label{M-Dilaton-r}{\cal M}_{a>1/3}&\sim& E^{\frac{2}{3a+1}},\\
       \label{M-Dilaton-ar}{\cal M}_{a\leq -1/3}&\sim &E^{\frac{2}{3(1-a)}}.\eea

Later, in \cite{Aref'eva:2014sua} we have confirmed   results  (\ref{M-Dilaton}), (\ref{M-Dilaton-a}) considering the domain wall collision models that generalized the
Lin-Shuryak model \cite{Lin,ABP} to non-conformal cases.
 In \cite{Aref'eva:2014sua}  we have also  noticed  that the model with the $b$-factor $b(z)=\frac{L_{eff}}{z}$
  gives a more realistic bound
  \be
  \label{M-ph-dilaton}{\cal M}_{ph-dilaton}(E)\sim E^{1/3}\ee
that is closer to (\ref{M-exp}).
But the price for this  modification is  the phantom  kinetic term for the dilaton.
 Note that we have not performed any UV cut-off  in this model to get estimation (\ref{M-ph-dilaton}).
\subsection{Trapped Surface for Domain-wall Shock Waves }
The equation for the domain-wall wave profile $\phi^w(z)$ in the space with the $b$-factor is
\be
\label{eq31}
\left(\partial^2_z+\frac{3b^\prime}{b}\partial_z\right)\phi^w(z)=-C\frac{\delta(z-z_*)}{b^3(z)},\ee
where   $C$ is a dimensionless variable
\be
\label{C-E}
C=\frac{16\pi G_5 E}{L^2}.\ee
The solution of (\ref{eq31}) is given as
\be
\phi^w(z)=\phi_a\Theta(z_*-z)+\phi_b\Theta(z-z_*),
\ee
where
\bea
\phi_a=C_a\int_{z_a}^z b^{-3}dz,\,\,\,\,\,
\phi_b=C_b\int_{z_b}^z b^{-3}dz.
\eea
The constants $C_a$ and $C_b$ can be represented in the form, see \cite{Lin,Aref'eva:2014sua}:
\bea
\label{STAR}C_a=C\frac{\int_{z_b}^{z_*} b^{-3}dz}{\int
_{z_b}^{z_a} b^{-3}dz},\,\,\,\,\,
C_b=C\frac{\int_{z_a}^{z_*} b^{-3}dz}{\int_{z_b}^{z_a} b^{-3}dz}.
\eea
As has been mentioned in Introduction we consider the collision of two shock domain walls in 5-dimensional space time as a holographical model of heavy ion collisions in real 4-dimensional space time. The shock wave profile  $\phi ^w$ satisfies equation (\ref{eq31}). The trapped surface formed in the wall-on-wall collision obeys equation (\ref{eq31}) and special boundary conditions. From these boundary conditions we can find that the trapped surface is located in z-direction from some point $z_a$ to point $z_b$, such that
$z_a<z_*<z_b$. The points  $z_a$ and $z_b$ can be found from the following relations
\bea
\label{star}\frac{C}{2}b^{-3}(z_a)\frac{\int_{z_b}^{z_*} b^{-3}dz}{\int^{z_a}_{z_b} b^{-3}dz}=1,\,\,\,\,\,
\frac{C}{2}b^{-3}(z_b)\frac{\int_{z_a}^{z_*} b^{-3}dz}{\int_{z_b}^{z_a} b^{-3}dz}=-1.
\eea
 Relations  (\ref{star}) guaranty that the trapped surface forms
and the trapped surface is located between $z_a $ and $z_b$ and the collision point $z_*$ is located
between $z_a $ and $z_b$, $z_a <z_*<z_b$  (see details in \cite{Lin,Aref'eva:2014sua}).

From (\ref{star}) we get
\bea
\label{C-b}
\frac{C}{2}=b^{3}(z_a)+b^{3}(z_b),\eea
\bea
F(z_*)=\frac{b^{-3}(z_a)F(z_b)+b^{-3}(z_b)F(z_a)}{b^{-3}(z_a)+b^{-3}(z_b)},
\eea
where
\be
\int_{z_i}^{z_j}b^{-3}dz=F(z_j)-F(z_i)\ee
and
\be
z_a<z_*<z_b.
\label{order}\ee

There is  the following formula for the entropy density \cite{Aref'eva:2014sua}
of the trapped surface
 \be\label{gen-entr}
 s=\frac{S_{\text{trap}}}{\int d^2x_{\perp}}=\frac{1}{2G_5}\int^{z_b}_{z_a}b^{3}\,dz.
 \ee

\section{Intermediate Background}
\subsection{Entropy}
In this section we consider the metric (\ref{metric-b}) with $b$-factor
\be
\label{b1}b=b_1\equiv \left(\frac{L_{eff}}{z}\right)^{1/2}.
\ee
The entropy dependence on the energy can be read from the formula
\be
\label{s1}
s_1=\frac{L_{eff}}{G_5}\left( \left(\frac{L_{eff}}{z_a}\right)^{1/2}-\left(\frac{L_{eff}}{z_b}\right)^{1/2}\right),
\ee
 where
\be
 \label{za-zb-1}
 \frac{z_a}{z_b}=\left( \frac{C}{2} \left(\frac{z_b}{L_{eff}}\right)^{3/2}-1\right)^{-2/3}.\ee
 Substituting (\ref{za-zb-1}) into (\ref{s1}) we get
 \bea\label{s1C}
 s_1(C,z_b)
&=& \frac{L_{eff}}{G_5}\left(\frac{L_{eff}}{z_b}\right)^{1/2}\left[ \left( \frac{C}{2} \left(\frac{z_b}{L_{eff}}\right)^{3/2}-1\right)^{1/3}- 1\right].\eea

\begin{figure}[h!]
    \centering
  \includegraphics[width=6cm]{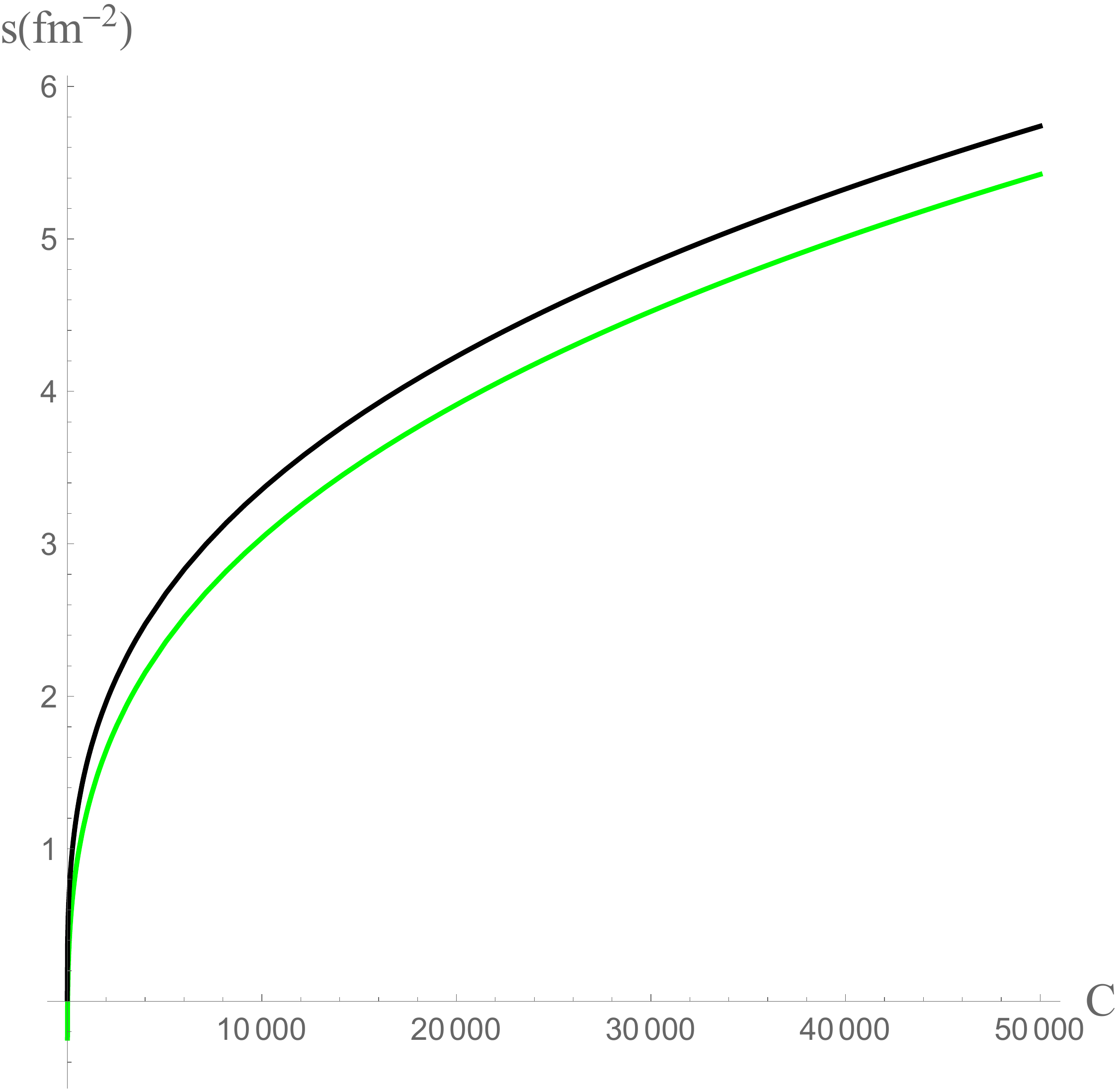} A.$\,\,\,\,$
  \includegraphics[width=6cm]{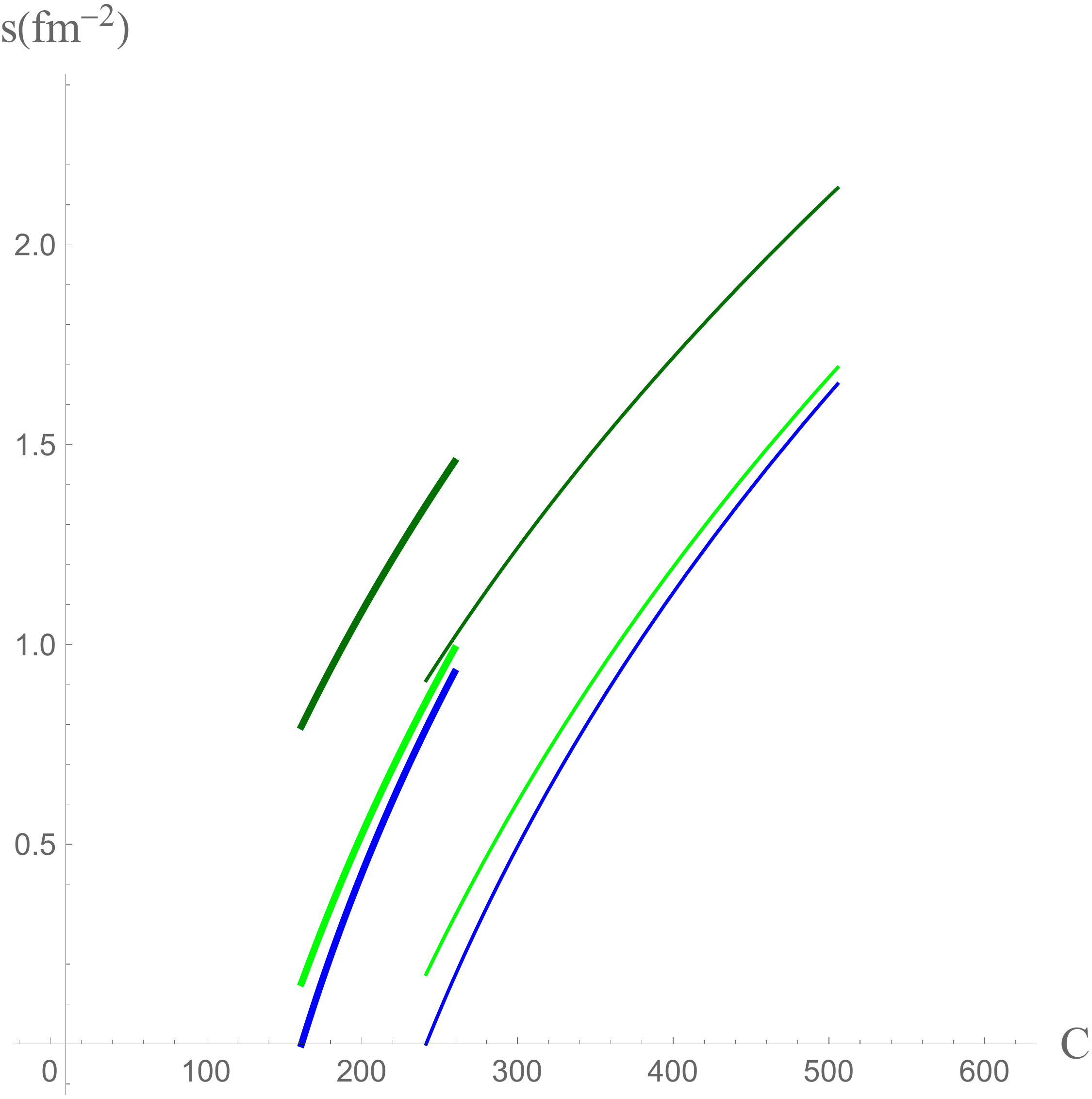} B.
      \caption{A. The black line represents $s^{(0)}_1(C)$ and the green one represents  $s_1(C,z_b)$  at $z_b=1.7 \, fm$.  Here we take $L_{eff}=4.4  \, fm$ and $G_5=44.83\,fm^3$.  B.
     The dependence of the entropy $s_1(C,z_b)$ 
    on $C$ for  $z_b$, $z_b=1.7 \, fm$ (blue thick line) and  $z_b=1.3 \, fm$ (blue  thin line).
    Approximations  $s^{(1)}_1(C,z_b)$ (khaki lines)  and  $s^{(2)}_1(C,z_b)$  (green lines)
      for  $z_b=1.7 \, fm$ (thick lines) and $z_b=1.3 \, fm$ (thin lines).  Here $L_{eff}=20.7 \, fm$ . For thick lines $z_a$ varies from $1 \, fm$ to $1.7 \, fm$, for thin lines $z_a$  varies from $0.6 \, fm$ to $1.3\, fm$.
          }\label{Fig:s1}
\end{figure}

One can perform the large $C$  expansion in formula (\ref{s1C}) to get
\bea
\label{s1-exp}
s_1(C,z_b)&=&\frac{L_{eff}}{G_5}\left(\left(\frac{C}{2}\right)^{1/3}-\left(\frac{L_{eff}}{z_b}\right)^{1/2}
-\frac{1}{3}\left(\frac{2}{C}\right)^{2/3}\left(\frac{L_{eff}}{z_b}\right)^{3/2}+...\right),\eea
i.e. the zeroth, first and  second approximations are given by
\bea
s^{(0)}_1(C)&=&\frac{L_{eff}}{G_5}\left(\frac{C}{2}\right)^{1/3},\\\nn
s^{(1)}_1(C,z_b)&=&\frac{L_{eff}}{G_5}\left(\left(\frac{C}{2}\right)^{1/3}-\left(\frac{L_{eff}}{z_b}\right)^{1/2}
\right),\\\nn
\\
s^{(2)}_1(C,z_b)&=&\frac{L_{eff}}{G_5}\left(\left(\frac{C}{2}\right)^{1/3}-\left(\frac{L_{eff}}{z_b}\right)^{1/2}
-\frac{1}{3}\left(\frac{2}{C}\right)^{2/3}\left(\frac{L_{eff}}{z_b}\right)^{3/2}\right).
\eea
The dependence of entropy $s_1(C,z_b)$ on  $C$   at fixed $z_b$  is
presented by the green line in Fig.\ref{Fig:s1}.A.  The  approximation $s^{(0)}_1(C)$ at large $C$ has behavior $C^{1/3}$ and it is shown by the black line in Fig.\ref{Fig:s1}.A. Note that  due to relation (\ref{C-E}) $C\sim E$.
More precisely, taking $G_5=44.83\,fm^3$ and $L_{eff}=4.4 \,fm$ we get $C=580 E/GeV$ and the range of variation $C$
at Fig.\ref{Fig:s1}.A corresponds to the energy around  0.1 TeV.
Therefore, one can say that at large energies $s_1(E,z_b)\sim E^{1/3}$  that in fact is rather close to the experimental dependence $\sim E^{0.3}$.

However  relation (\ref{za-zb-1}) written in the form 
\be
 \label{za-zb-1-m}
 \left( \frac{L_{eff}}{z_a}\right)^{3/2}+ \left( \frac{L_{eff}}{z_b}\right)^{3/2}=\frac{C}{2}.\ee
 may give  a restriction on possible  variation range of energy. Indeed,
 by the construction $z_a<z_b$ and to get a large value of $C$ at fixed $z_b$ one has to take a small $z_a$.
 In the case of an additional restriction on $z_a$, say $z_a>z_{a, min}$, we get a restriction $C<C_{max}$.  It may happen that in this area of $C$ we cannot restrict ourself by the zeroth approximation
 to $s_1(C,z_b)$.  Few examples of such behavior are presented in  Fig.\ref{Fig:s1}.B.
In Fig.\ref{Fig:s1}.B the energy dependencies of the exact entropy $s_1(C,z_b)$ and approximated entropies
$s^{(1)}_1(C,z_b)$ and $s^{(2)}_1(C,z_b)$ for different $z_b$  are shown for relatively small values of $C$.
Fig.\ref{Fig:s1}.B shows that  in the considered regions of $C$ we have to take into account the first three terms
of the approximation (\ref{s1-exp}). The choice  of $L_{eff}=20.7 \, fm$ in Fig.\ref{Fig:s1}.B will be claire in Sect.4.

\subsection{Thermalization Times}
\begin{figure}[h!]
    \centering
     \includegraphics[width=7cm]{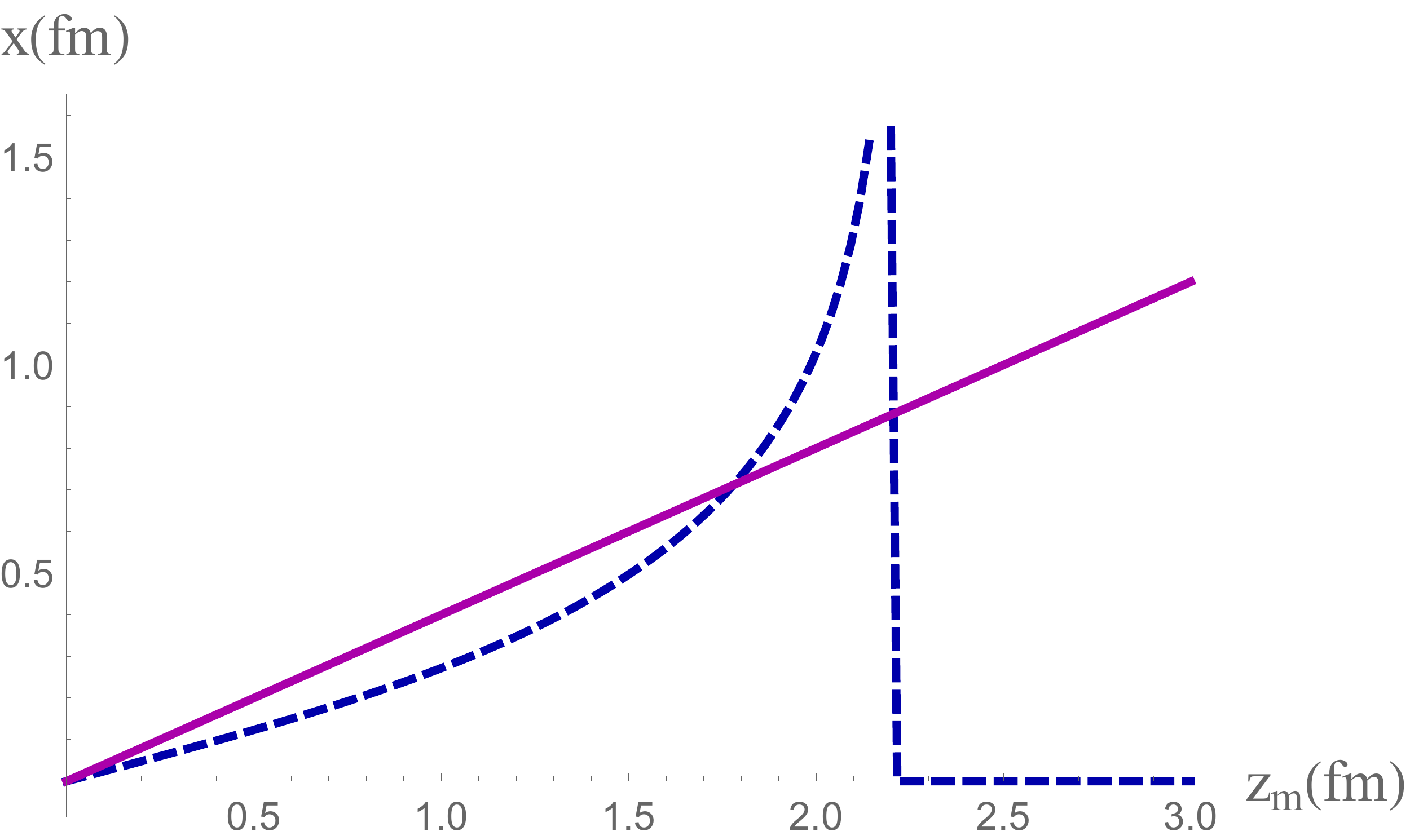}
     \caption{ Dependence of the interquark distance  $x$  on the string maximum holographic coordinate $z_m$ for metric with the factor $b_1(z)$ (solid magenta line) and for the metric $b_2(z)$ (dashed blue line). }
     \label{Fig:b1x-zm}
\end{figure}

We estimate the thermalization time  by a characteristic size of the trapped surface, i.e.
\be
\label{timetherm}
\tau_{therm}\sim \frac{z_b-z_a}{2.4}.\ee
We put the factor 2.4   taking into account the relation between the interquark distance  $x$  and
the string maximum holographic coordinate $z_{m}$.  The dependence of the interquark distance $x$
 on the maximum of  string  $z$-coordinate,  $z_m$,  is given by
\be
\label{x-zm}
x=\int_{0}^{z_m} \frac{2}{\sqrt{\displaystyle\frac{b^4(z)}{b^4(z_m)}-1}}{dz},
\ee
see for details for example \cite{{Pirner:2009gr}}. For metric
(\ref{metric-b}) with $b_1$ given by (\ref{b1})  this dependence is presented in Fig.\ref{Fig:b1x-zm}
by the solid magenta line and we see that $x=z_m/2.4$.

Note that the formula (\ref{timetherm}) is written using general causal arguments.
We assume that the time of formation of an object
extended along the holographic direction from  $z_a$  to $z_b$ is the same as the time
formation of an extended object along x-direction with a  characteristic scale  $\Delta x=(z_b-z_a)/2.4$. This is  in
accordance with  (\ref {x-zm}). Formation  of such object
can be performed no faster as $\Delta x$.

\begin{figure}[h!]
    \centering
        \includegraphics[width=7cm]{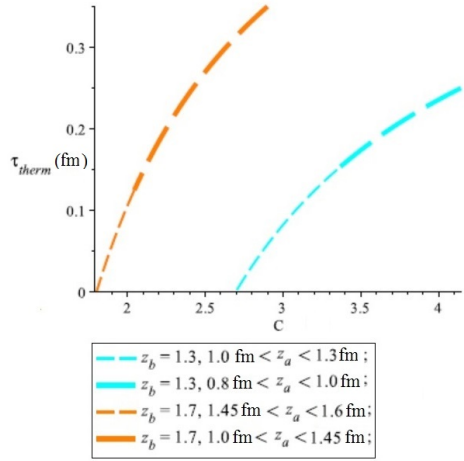}
     \caption{The dependence of the thermalization time on  $C$ for different values of $z_b$  ($z_b =1.7 \, fm$
      -- coral lines, $z_b =1.3 \, fm$ -- cyan lines).  Different  range of variation  of  $z_a$ are shown by lines with different thickness.}
     \label{Fig:b1x-z}
\end{figure}

The dependence of the thermalization time on  $C$ for a given value of $z_b$ can be estimated substituting  $z_a$ from
(\ref{za-zb-1}) into the right hand site of  (\ref{timetherm}).
In Fig.\ref{Fig:b1x-z}  the dependence of the thermalization time on  $C$ for different values of $z_b$ is presented.
To vary $C$ we vary $z_a$.
In the plot different  domains of  $z_a$ are shown by lines with different thickness.
 Small values of $z_a$
correspond to large values of the energy.

\newpage
$$\,$$
\newpage
\section{Intermediate Background as a Part of Confining Background}
\begin{figure}[h!]
    \centering
     \includegraphics[width=10cm]{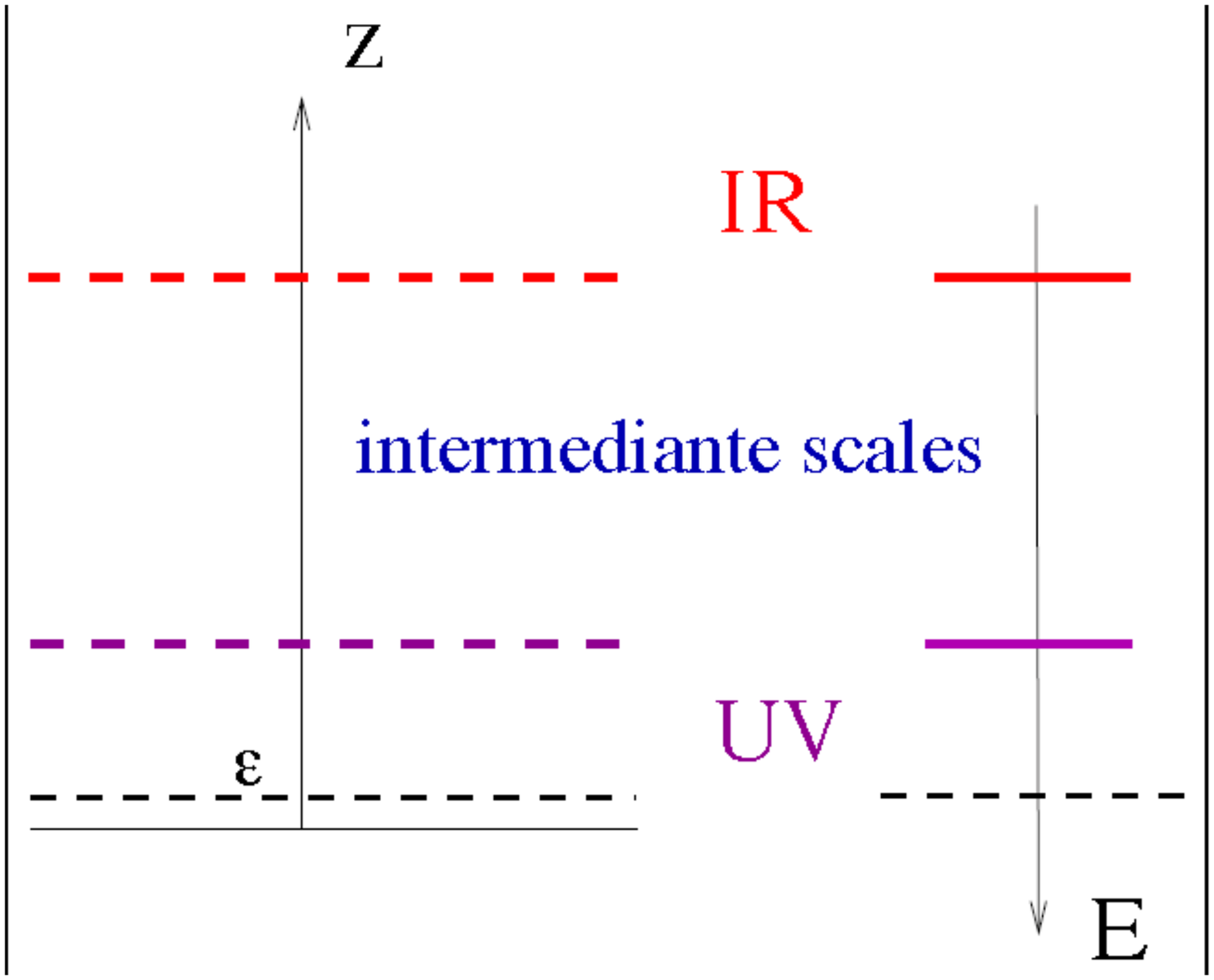}
        \caption{Schematic picture of the bulk scales.}\label{Fig:b1b2-mmm}
\end{figure}

In this section we consider the metric (\ref{metric-b})  with confining factor $b(z)$
\be
\label{b2}
b(z)=b_2(z)\equiv\frac{Le^{\frac{az^2}{4}}\sqrt{1+gz}}{z},\,\,\,\,\,\,\,\,\,g=-0.02\,GeV. \ee
A
 schematic picture of the bulk scales is presented in Fig.\ref{Fig:b1b2-mmm}.

\begin{figure}[h!]
    \centering
      \includegraphics[width=7cm]{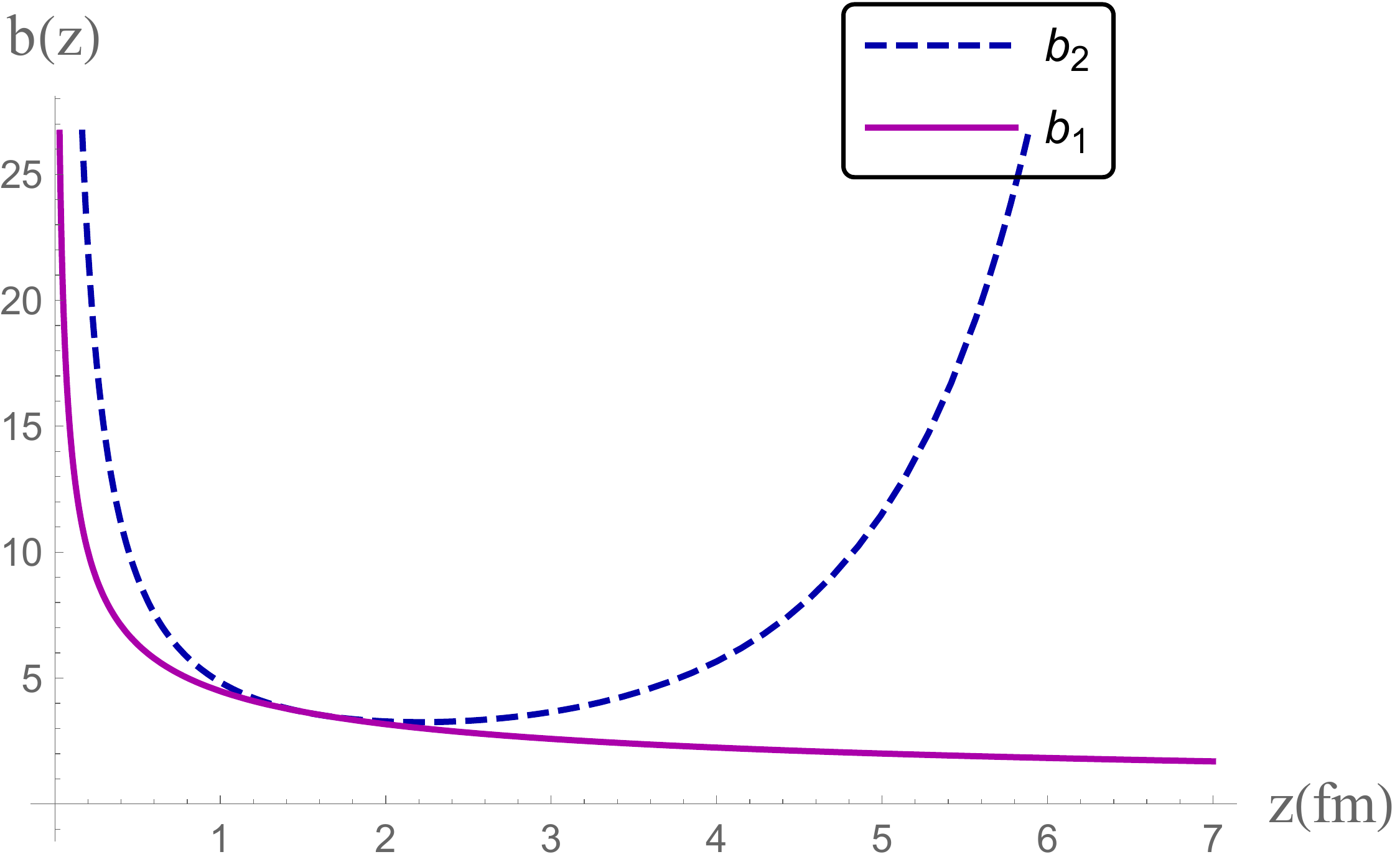} A.
     \includegraphics[width=7cm]{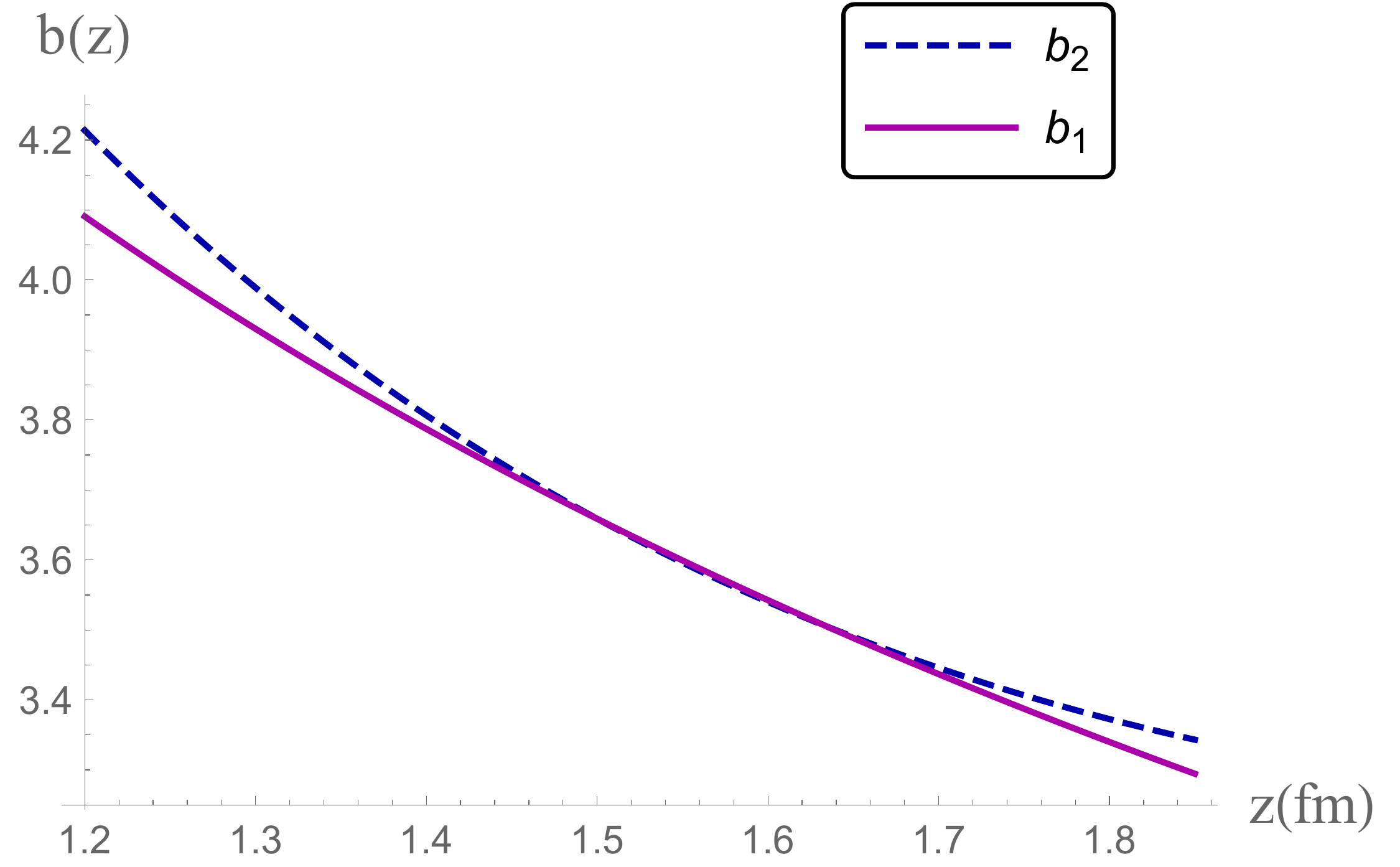}\,\,\,\,\,B.
        \caption{A. $b$-factors $b_1(z)$ and $b_2(z)$.  Here  $L=4.4 \, fm$ and $L_{eff}=20.86 \, fm$. B.
     The same $b$-factors as in A in the  intermediate region $1.2 \, fm<z <1.8 \, fm$.
      Solid magenta lines correspond to $b_1=b_1(z)$, dashed blue lines correspond to $b_2=b_2(z).$
     }\label{Fig:b1b2-m}
\end{figure}

\begin{figure}[h!]
    \centering
     \includegraphics[width=7cm]{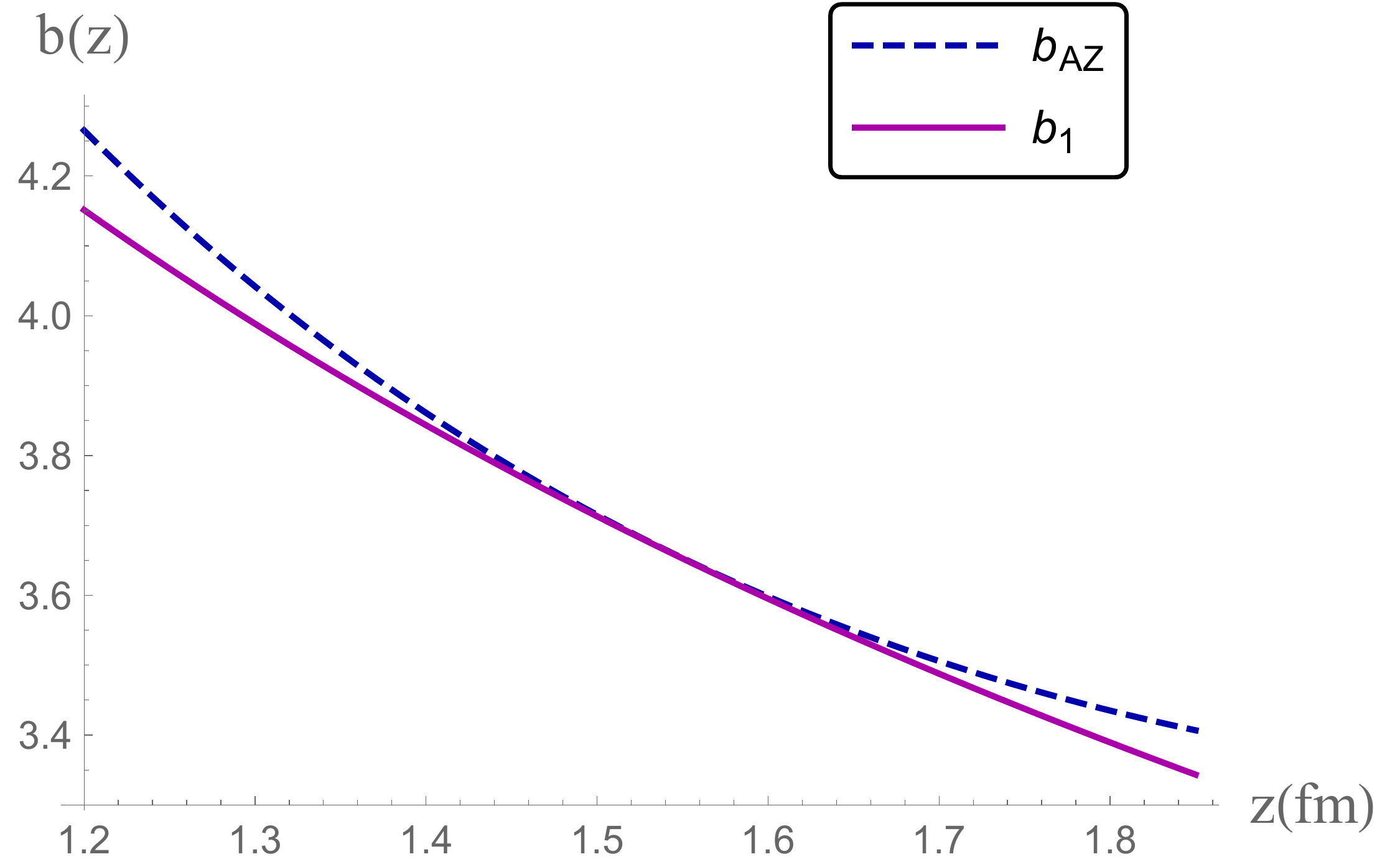}
        \caption{
     Factors $b_1(z)$ with $L_{eff}=20.7 \, fm$ (solid magenta line) and $b(z)$  given by (2.2) (dashed blue line) in the  intermediate region $1.2 \, fm<z <1.8 \, fm$.
   }\label{Fig:AZ}
\end{figure}

 Fig.\ref{Fig:b1b2-m} shows that for $L=4.4fm$ at the region of intermediate  holographic  coordinate $z$,
$1.3 \, fm=z_{UV}<z<z_{IR}=1.8 \, fm$  the factor  $b_1$ with  $L_{eff}=20.86\, fm$ coincides with  $b_2$  up to 3\%  and  for $1.4 \, fm <z<1.7 \, fm$
these factors are almost coincided. Instead of \eqref{b2} we can use the $b$-factor \eqref{b-AZ} from \cite{Andreev:2006ct}. This leads to
a slight variation of parameters $L_{eff}$, $z_{IR}$ and $z_{UV}$, see Fig.\ref{Fig:AZ}.

Note that the metric with $b$-factor $b_1$ leads to the potential of interquark interaction of the form $V(r)\sim A \log{(\frac{r}{r_0})}$, where $A$ and $r_0$ are some constants. This form of the potential was suggested as the simple model to fit identical spin-averaged charmonia and
bottomonia level splitting, see \cite{Bali:2000gf} and refs therein.

The dependence of the interquark distance $x$
 on the string maximum   $z$-coordinate $z_m$ for metric (\ref{metric-b})  with confining factor $b_2$ is presented in
Fig.\ref {Fig:b1x-zm}  by dashed blue line. Note that  at $z\approx2.2 \, fm$
there is the string breaking, that is in accordance  with \cite{stringbreaking}.  This point is out of our intermediate zone.

In formula for entropy (\ref{gen-entr}) we take the usually accepted value $G_5\approx 44.83 \, fm^3 $ \cite{Gubser}.


\begin{figure}[h!]
    \centering
     \includegraphics[width=6cm]{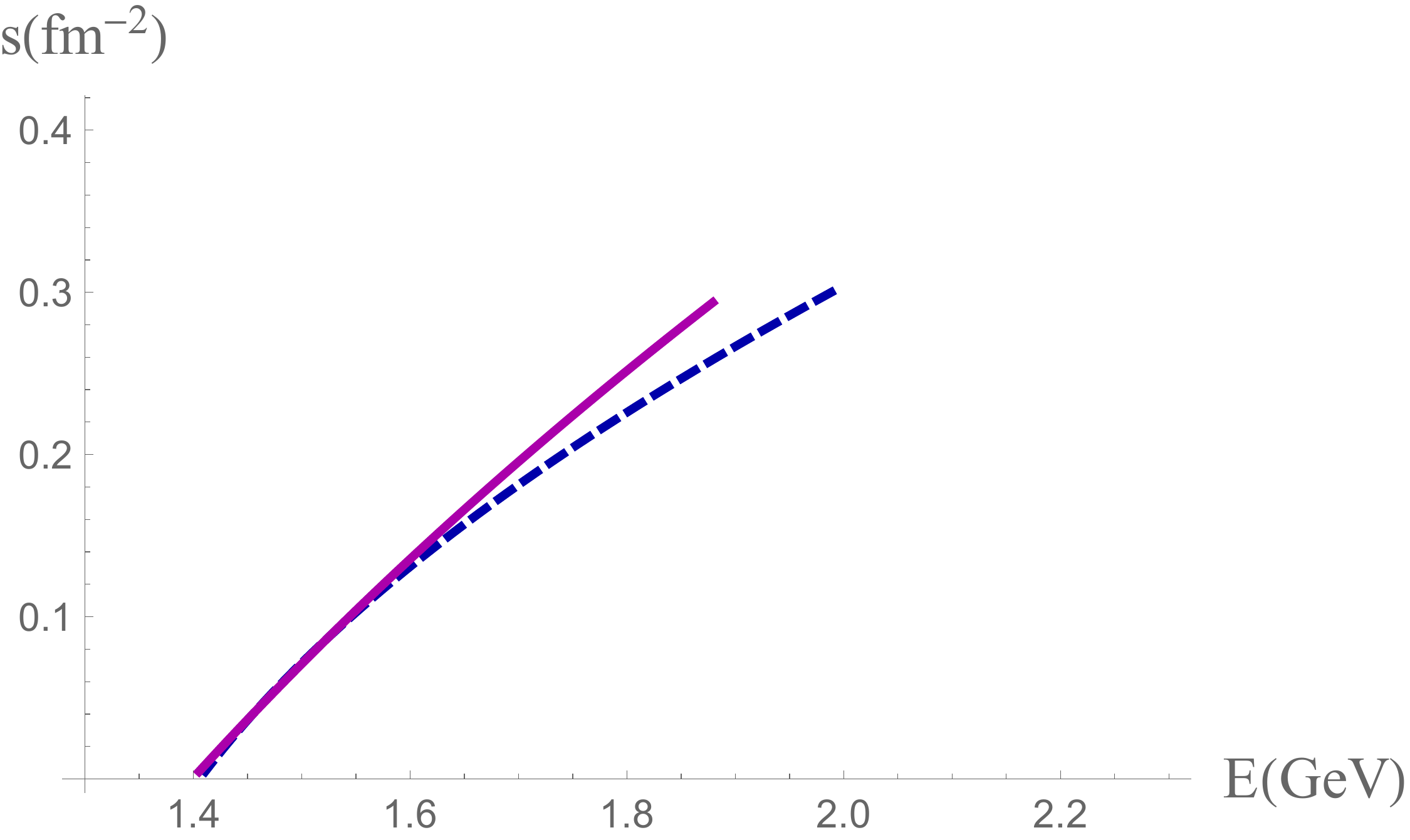}\,\,\,A.\,\,\,\,\,\,\,\,
       \includegraphics[width=6cm]{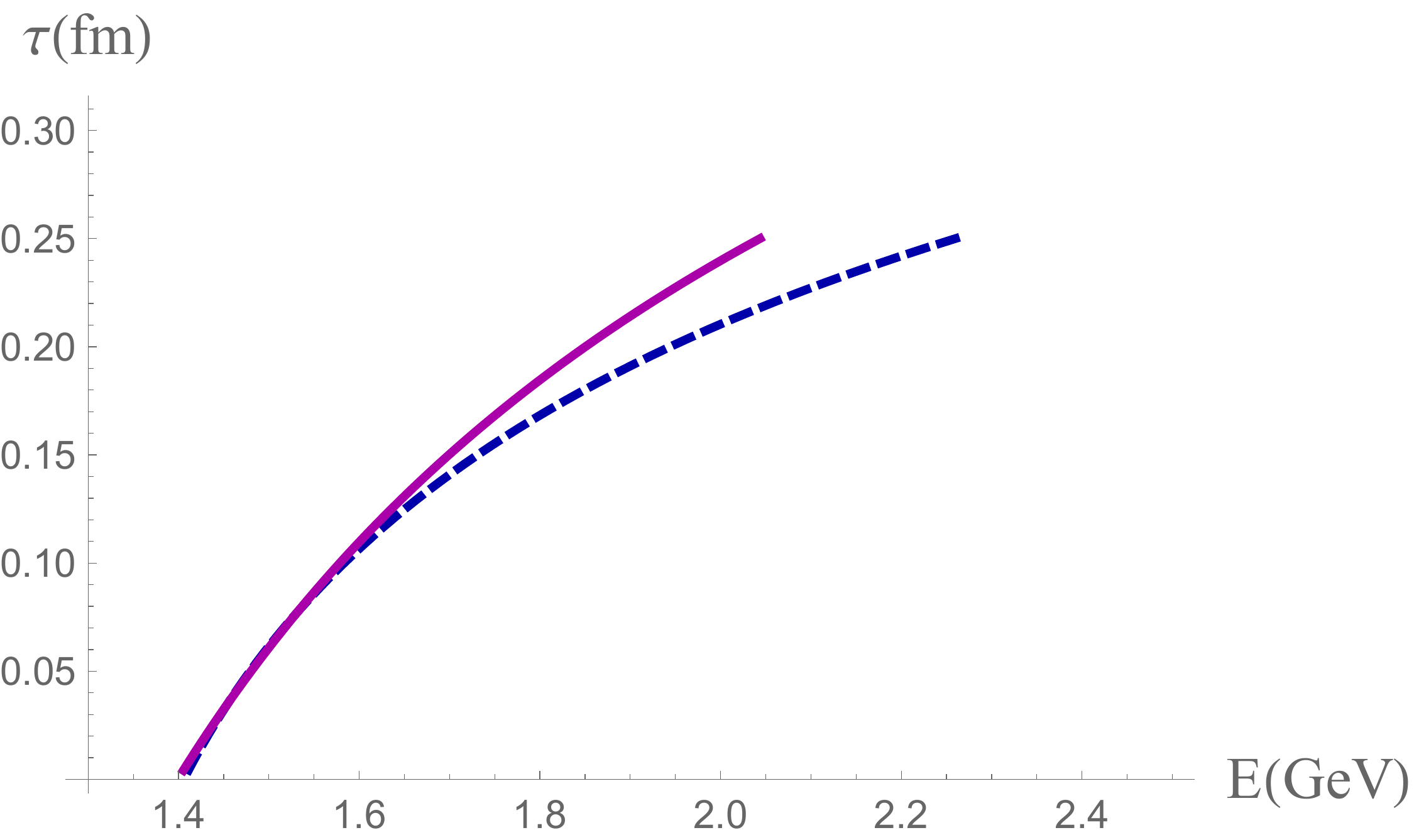}\,\,\,\,\,B.
           \caption{A. The entropy dependence on the energy for the confining metric (dashed line)
           and
            $b_{1}^2=\frac{L_{eff}}{z}$ (solid line). For both lines $z_b=1.8 \, fm$ and $z_a$ varies from $z_a=1.2\,fm$ to  $z_a=1.8 \, fm$.
B. The thermalization time dependence on the energy. The dashed line corresponds to the confining metric, the solid line corresponds to $b_{1}^2=\frac{L_{eff}}{z}$. $z_b$ and $z_a$ are the same as in A. 
        }\label{Fig:s_b1b2}
\end{figure}

In Fig.\ref{Fig:s_b1b2} A. the entropy dependence on the energy is presented
for the confining metric (dashed line)
           and
            $b_{1}^2=\frac{L_{eff}}{z}$ (solid line). For both lines $z_b=1.8 \, fm$ and $z_a$ varies from $z_a=1.2\,fm$ to  $z_a=1.8 \, fm$.
  We see that the energy dependence of entropies are very close
in this  intermediate region $1.2 \, fm<z <1.8 \, fm$.
This intermediate region according (\ref{timetherm}) corresponds to the thermalization time $\tau_{therm}\approx 0.25\, fm.$

Let us note, that our assumption about restriction of area of the trapped surface formation and consideration here
instead  of the confining metric with $b$-factor (\ref{b2}) the non-confining  one with $b$-factor  (\ref{b1}),  that in the asymptotic regime gives the desirable energy dependence of entropy, is  similar in some sense to a proposal to use
the energy-dependent cut-off in the high-energy limit \cite{Gubser2}.  We can also say that our estimations
  give an analytical realization of this proposal.

  However, if we consider a wider region for possible values of $z_a$  and consider for example $z_a\to 0$, that corresponds to
 large energies, we get
   different entropy behavior in these two models.  The model with factor $b_2$  has a typical behaviour
  $s(C)\sim C^{2/3}$ \cite{Gubser,ABP} and the model with factor $b_1$ has   $s(C)\sim C^{1/3}$ at large $C$ \cite{APP}.
  This means that without change of asymptotics of the  b-factor at the UV region we cannot change behavior of $s(C)$
  at large energies. From other point of view the UV asymptotics of the b-factor is fixed by the Coulomb  potential \cite{Malda}.
  This lead us to  a modification  of the  holographic scenario and consideration at small $z$ an anisotropic background, where we can expect that the most part of the entropy for large energy is produced \cite{IAAG}.

\section{Conclusion}

Our calculations show that,  within the holographic model of heavy-ions collisions using the confining
vacuum background and colliding domain shock waves, the produced entropy   has an asymptotic expansion,
the first term of which  provides  a suitable dependence  on the energy $s_2^{(0)}\sim E^{1/3}$.
However, the entropy produced during a  time  $\sim 0.25 fm$ after colliding of two shock domain walls in the confining background cannot be saturated by the first term  and  contributions of non-leading terms 
have to be taken into account.  
This is related with the fact that to restrict our asymptotic expansion by the first term we   have  to consider 
the asymptotic expansion at large energies. Large energies correspond to small values of the holographic 
coordinate $z$, where our approximation of the metric with $b_2$ factor (\ref{b2}) by the  metric with $b_1$ factor (\ref{b1}) 
fails. From the other hand, as mentioned in the text, we cannot change the asymptotic of $b_2$ since it is related with the Coulomb potential. 

It seems that one of possible resolutions of the problem is a change of the scenario of the isotropic holographic  thermalization
to   an anisotropic short time  holographic  thermalization scenario. This scenario  assumes that 
the main part of multiplicity is produced in an anisotropic regime  and this part of multiplicity  
can be estimated by the trapped surface  produced under a collision of the two shock waves in an anisotropic background. This scenario is accepted in the recent paper \cite{IAAG}, where collisions of shock waves in the Lifshitz-like background  have been considered. 

It would be interesting to compare
our estimation of the thermalization time with thermalization time estimations given by the Vaidya
confining bulk metric, as well the thermalization time obtained in holographic hard wall model using the homogeneous injection of the energy \cite{1311.7560}.

  Note, that our consideration may have applications not only for heavy-ions collisions, but also in studies of
 thermalization process in a broader class of
 strongly correlated multi-particle  systems.

\section*{Acknowledgments}
It is our pleasure to dedicate this paper to academician Valery Rubakov on the occasion of his 60th birthday.
This work was  supported by the Russian Science Foundation  grant 14-11-00687. We are grateful to Oleg Andreev for useful remarks about the first version of our paper.

\end{document}